\documentclass[tightenlines,twoside,secnumarabic,
               onecolumn,floatfix,nofootinbib,showpacs,11pt]{revtex4}

\usepackage[english]{babel}
\usepackage{amsmath,amssymb}
\usepackage{graphicx}
\usepackage[sort&compress]{natbib}

\allowdisplaybreaks

\bibpunct{[}{]}{,}{n}{}{,}

\newcommand{\iso}[2]{{\ensuremath{{}^{#2}}\ensuremath{\rm #1}}}
\newcommand{\bb}{{\ensuremath{\text{b}\beta}}}
\newcommand{\cb}{{\ensuremath{\text{c}\beta}}}

\begin{document}

\title{Ultra-low $Q$ values for neutrino mass measurements}
\author{Joachim Kopp$^{1,2}$}    \email[Email: ]{jkopp@fnal.gov}
\author{Alexander Merle$^1$} \email[Email: ]{amerle@mpi-hd.mpg.de}
\affiliation{$^1$ Max-Planck-Institut f\"ur Kernphysik,
             Postfach 10 39 80, 69029 Heidelberg, Germany \\
             $^2$ Fermilab, P.O.~Box 500, Batavia, IL 60510-0500, USA}
\date{November 17, 2009} 

\begin{abstract}
  We investigate weak nuclear decays with extremely small kinetic energy
  release ($Q$ value) and thus extremely good sensitivity to the absolute
  neutrino mass scale. In particular, we consider decays into excited daughter
  states, and we show that partial ionization of the parent atom can help to
  tune $Q$ values to $\ll 1$~keV. We discuss several candidate isotopes
  undergoing $\beta^\pm$, bound state $\beta$, or electron capture decay, and
  come to the conclusion that a neutrino mass measurement using low-$Q$ decays
  might only be feasible if no ionization is required, and if future
  improvements in isotope production technology, nuclear mass spectroscopy, and
  atomic structure calculations are possible. Experiments using ions, however,
  are extremely challenging due to the large number of ions that must be
  stored. New precision data on nuclear excitation levels could help to
  identify further isotopes with low-$Q$ decay modes and possibly less
  challenging requirements.
\end{abstract}

\begin{flushright}
  FERMILAB-PUB-09-582-T
\end{flushright}

\maketitle

\section{Introduction}
\label{sec:intro}

One of the big unknowns in astroparticle physics today is the absolute neutrino
mass scale $m_\nu$. While indirect probes such as
cosmology~\cite{Elgaroy:2004rc} and neutrinoless double $\beta$
decay~\cite{Avignone:2007fu} achieve sub-eV sensitivity to $m_\nu$,
it is desirable to complement these measurements with model-independent direct
bounds. The most advanced efforts in this direction have been the kinematical
studies of the $\beta$ spectrum in tritium decay by the
Mainz~\cite{Kraus:2004zw} and Troitsk~\cite{Belesev:1995sb} collaborations,
yielding the limit $m_\nu \lesssim 2$~eV. In the near future, the sensitivity
will be improved to $m_\nu \lesssim 0.2$~eV by the KATRIN
experiment~\cite{Angrik:2005ep}.  However, Mainz, Troitsk, and KATRIN are
limited by the accuracy to which the spectrum of decay electrons can be
measured few eV below the kinematical endpoint, where the impact of $m_\nu>0$
is largest.  Since the kinetic energy release ($Q$ value) of tritium decay is
$18.6$~keV, only a very small fraction of decays falls into that region so that
large statistics, very good background suppression, and an excellent energy
resolution are required. If KATRIN should not see a positive signal, new
experimental techniques would be required to push the sensitivity to even
smaller $m_\nu$. For example, it has been proposed to study nuclear recoils in
bound state $\beta$ decay of tritium~\cite{Cohen:1987}, to reconstruct the
electron and nuclear kinematics in tritium decay~\cite{Jerkins:2009tc}, or to
measure the electron flux near the tritium endpoint in a storage
ring~\cite{Lindroos:2009mx}.  However, all of these proposals are limited by
the large $Q$ value of tritium, which makes the neutrino mass a small effect.
The decay $\iso{Re}{187} \rightarrow \iso{Os}{187}$ offers a lower $Q$ value of
only $2.657$~keV and thus better sensitivity to $m_\nu$, but since it is a
unique first forbidden decay, the small decay rate makes it difficult to
accumulate sufficient statistics~\cite{Monfardini:2005dk}.

\begin{figure}
  \vspace{-.3cm}
  \begin{center}
    \includegraphics[width=9cm]{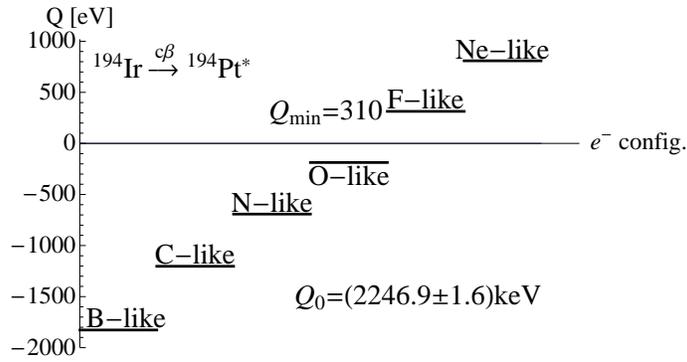}
  \end{center}
  \vspace{-.5cm}
  \caption{An example for tuning of a $Q$ value by ionization. Note that energy
    levels in the decay $\iso{Ir}{194} \to \iso{Pt^*}{194}$ shown here have an
    uncertainty of $\mathcal{O}(1.6\ \textrm{keV})$, so the figure is to be
    understood as an illustration of the principle idea only.}
  \label{fig:Qvalues}
\end{figure}

In this paper, we investigate weak decays with even smaller $Q$ values. In
particular, we consider continuum $\beta$ ($\cb$), bound state $\beta$ ($\bb$),
and electron capture (EC) decays. The key ideas are to consider decays to
\emph{excited} nuclear daughter states and to use \emph{ions} instead of
neutral atoms if necessary.
As illustrated in Fig.~\ref{fig:Qvalues}, an
appropriate choice of the ionization level allows for some tuning of the $Q$
value since every spectator electron contributes to $Q$ with its energy gain or
loss due to the change of the nuclear charge during the decay. For $\bb$
decay~\cite{Bahcall:1961,Jung:1992pw}, ionization can also have the direct
effect of opening up new decay modes.  Our aim is to find decays that have
sufficiently small $Q$ values to depend appreciably on $m_\nu$, but at the same
time still have an absolute rate large enough to allow for a good
signal-to-noise ratio.

For $\bb$ and EC decay, the observable sensitive to $m_\nu$ is the decay rate,
which could be measured by detecting gamma and x-ray photons accompanying the
decay. In the case of $\cb$ decay, the sensitivity can be increased by guiding
the decay electrons into a spectrometer similar to the ones used in Mainz,
Troitsk, and KATRIN in order to also measure the $\beta$ spectrum near the
endpoint. For those decays where ionization is required to achieve sufficiently
low $Q$, we propose to store the parent ions in a trap or in a storage ring. We
discuss the feasibility of these ideas below.

\section{Nuclear decays with ultra-low $Q$ values}
\label{sec:examples}

Nuclear decays with an ultra-small kinetic energy release $Q \ll 1$~keV can
occur only if the daughter nucleus has a state with excitation energy $E^*\geq0$
fulfilling
\begin{alignat*}{2}
 &\textrm{$\beta^-$:}&\quad
   Q_0 - (B_{Z+1,Z} - B_{Z,Z})  &\lesssim  E^*
     \lesssim  Q_0 + B_{Z,1} - B_{Z+1,2} \,,  \\
 &\textrm{$\beta^+$:}&\quad
   Q_0 - 2 m_e   &\lesssim  E^*  \lesssim  Q_0 - 2 m_e + B_{Z,Z} - B_{Z-1,Z} \,,  \\
 &\textrm{EC:}&\quad
   Q_0 - B_{Z,1} &\lesssim  E^*  \lesssim Q_0 \,.
\end{alignat*}
Here, we have neglected contributions from the $\mathcal{O}({\rm eV})$ binding
energies of outer shell electrons. $Q_0$ refers to the atomic mass difference
of the parent and daughter nuclei\footnote{Note that in most reference tables,
the atomic mass difference is simply called $Q$, while we reserve that notation
for the actual kinetic energy release in a decay, which is the quantity that
determines the sensitivity to $m_\nu$.} and $B_{Z,n}$ is the modulus of the
total electron binding energy in an atom or ion with nuclear charge $Z$ and
$n$ orbital electrons. In the first equation, $(B_{Z+1,Z} - B_{Z,Z})$ is the
energy gain of the spectator electrons in the decay of a neutral atom. By
ionization, the effective $Q$ value can be reduced by up to that amount.
In \bb\ decay, ionization can also increase $Q$ by opening up decay modes to
low-lying bound states. The maximum possible increase occurs for \bb\ decay of
a hydrogen-like ion into a helium-like daughter state, and is consequently of
$\mathcal{O}(B_{Z,1} - B_{Z+1,2})$. Similarly, for $\beta^+$ decay, $Q$ can be
increased compared to $(Q_0 - 2 m_e)$, the value for neutral atoms, by removing
spectator electrons and thus avoiding an energy loss of up to $(B_{Z,Z} -
B_{Z-1,Z})$. For EC, $Q$ can be smaller than the atomic mass difference $Q_0$
by up to the binding energy of the $1s$ electrons, which is of
$\mathcal{O}(B_{Z,1})$. $Q$ cannot be made significantly larger than $Q_0$ for
EC.

We list several candidate isotopes for low-$Q$ $\cb^\pm$, $\bb^-$, and EC decay
in Table~\ref{tab:isotopes}. Since nuclear structure data is still very
incomplete for many isotopes, it is quite possible that other suitable decays
will be identified in the future. To keep the expected signal-to-background
ratio large, we have only considered isotopes for which a low-$Q$ decay is
allowed from spin and parity arguments, while other decay modes (if present)
are at least first forbidden or otherwise have a very small branching ratio.
Note that, for some of our candidate isotopes, decay into the relevant excited
daughter state $E^*$ has not been observed yet, so even though it is not
forbidden by spin and parity arguments we cannot be sure that it exists.  For
each decay we have computed $Q$ as a function of the electron configuration.
The main uncertainties in this calculation come from the atomic mass
differences $Q_0$, which are typically known to
$\mathcal{O}(\text{keV})$~\cite{Firestone}, and from the binding energies of
multi-electron configurations. We have estimated these binding energies using
(I) the relativistic Hartree-Fock code {\tt atsp2K}~\cite{FroeseFischer:1997,
atsp2k} and (II) published atomic physics data and simulation
results~\cite{Rodrigues:2004,Johnson:1985,Plante:1994,Bearden:1967a} (we only
report the results of method (II)).  The good agreement between the two
independent estimates (I) and (II) shows that the atomic physics uncertainty in
our $Q$ values is $\lesssim 100$~eV and thus smaller than the uncertainties in
most $Q_0$ values.  An actual neutrino mass measurement would, however, require
both, $Q_0$ and the electron binding energies, to be known to an accuracy better
than $\mathcal{O}(m_\nu)$, and we discuss below how this could be achieved.
Here, we deal with the uncertainties by reporting how small $Q$ can be made if
the present best fit values for $Q_0$ are taken at face value, and by how much
$Q$ can change if $Q_0$ is varied within present uncertainties.  In all cases,
we assume the degree of ionization and the daughter state $E^*$ to be chosen in
the optimum way.

For \cb\ decay, the most promising isotopes at present are \iso{W}{188},
\iso{Os}{193}, and \iso{Ir}{194} with achievable $Q$ values between 0 and
1.3~keV, depending on the true value of $Q_0$. A measure for the sensitivity of
these low-$Q$ \cb\ decays to nonzero $m_\nu$ is the rate of events with
electron energies in a small interval $[Q - \delta E, Q]$ near the spectral
endpoint. However, by considering the phase space factor and the Coulomb
correction term (Fermi function) entering in the $\cb$ decay rate, it is easy
to show that this number is \emph{independent} of $Q$. To zeroth order, and
neglecting differences in nuclear matrix elements, this seems to indicate that
for achieving the same sensitivity as KATRIN in a low-$Q$ experiment a similar
number of stored parent atoms ($10^{19}$) would be required, which is far
beyond the capabilities of present ion traps ($\lesssim
10^6$--$10^8$)~\cite{Ames:2005a,PodaderaAliseda:2006rx} and storage rings
($\lesssim 10^9$--$10^{11}$)~\cite{KalantarNayestanaki:2009a}. However, the
larger relative effect of $m_\nu$ makes a low-$Q$ \cb\ decay experiment more
robust against many systematic errors. For instance, the required relative
spectrometer resolution is smaller than in the tritium case. Moreover, if the
time of each decay can be tagged by observing an associated gamma or x-ray
photon, the spectrometer can be operated in the more sensitive time-of-flight
(MAC-E TOF) mode~\cite{Angrik:2005ep}. Finally, it might be possible to combine
our ideas with the methods proposed
in~\cite{Cohen:1987,Jerkins:2009tc,Lindroos:2009mx} to measure also the energy
and momentum of the recoil nucleus.  All these effects should help to reduce
the required number of stored ions, even though the experiment will still be
extremely challenging.

\begin{table*}
  \begin{ruledtabular}
  \begin{tabular}{r@{$\,\rightarrow\!\!$}lccccp{5.5cm}}
    \multicolumn{2}{l}{Decay}       & $t_{1/2}$ & $Q_0$~[keV]        & $E^*$~[keV] & $Q$ [eV]               & Comment\\
    \hline
    \multicolumn{7}{l}{\bf Continuum $\beta^-$ decay} \\
    \iso{W}{188}  & \iso{Re^*}{188} & 69.4~d    & $349 \pm 3$        &   346.58    & $80^{+150}_{-80}$      & decay to $E^*$ not yet observed \newline
                                                                                                              decay impossible for unfavorable $Q_0$ \newline
                                                                                                              daughter spin uncertain \\
    \iso{Os}{193} & \iso{Ir^*}{193} & 30.5~h    & $1140.6 \pm 2.4$   & 1,131.2     & $50^{+1150}_{-50}$     & decay to $E^*$ not yet observed \\
    \iso{Ir}{194} & \iso{Pt^*}{194} & 19.15~h   & $2246.9 \pm 1.6$   & 2,239.8     & $310^{+200}_{-310}$    & decay to $E^*$ not yet observed \\
    \hline                                                                                                     
    \multicolumn{7}{l}{\bf Bound state $\beta^-$ decay} \\                                                     
    \iso{Dy}{163} & \iso{Ho}{163}   & stable    & $-2.576 \pm 0.016$ &     0       & $\approx 1,500$        & \\
    \hline                                                                                                     
    \multicolumn{7}{l}{\bf Continuum $\beta^+$ decay} \\                                                       
    \iso{Pt}{189} & \iso{Ir^*}{189} & 10.87~h   & $1971 \pm 14$      &   958.6     & $1880^{+670}_{-1180}$  & allowed background modes \newline
                                                                                                              \hspace*{.7em} with \%-level branching ratio \newline
                                                                                                              decay impossible for unfavorable $Q_0$ \\
    \hline                                                                                                     
    \multicolumn{7}{l}{\bf Electron capture decay} \\                                                       
    \iso{Dy}{159} & \iso{Tb^*}{159} & 144.4~d   & $365.6 \pm 1.2$    &   363.51    & $130^{+1200}_{-130}$   & might not require ionization \\
    \iso{Ho}{163} & \iso{Dy}{163}   & 4570~y    & $-2.576 \pm 0.016$ &     0       & $\approx 540$          & might not require ionization
  \end{tabular}
  \end{ruledtabular}
  \vspace{-.2cm}
  \caption{Candidates for ultra-low $Q$ decays. The calculation of $Q$ values
    is based on data from~\cite{Firestone,Rodrigues:2004,Johnson:1985,Plante:1994,
    Bearden:1967a}.}
  \label{tab:isotopes}
\end{table*}

\begin{figure}
  \vspace{-.3cm}
  \begin{center}
    \includegraphics[width=7cm]{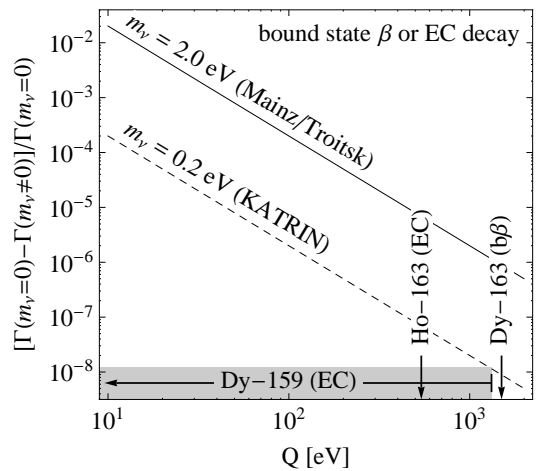}
  \end{center}
  \vspace{-.5cm}
  \caption{Relative effect of nonzero $m_\nu$ on the decay rate in \bb\ and EC decay.}
  \label{fig:bb}
\end{figure}

For $\bb$ decay, \iso{Dy}{163} could provide $Q \sim 1.5$~keV. This isotope has
the interesting property of being stable as a neutral atom, but becoming
unstable to \bb\ decay when ionized~\cite{Takahashi:1983a}.  The most promising
isotopes undergoing electron capture are $\iso{Dy}{159}$ and $\iso{Ho}{163}$,
for which $M$-capture with a very low $Q$ value might occur even without ionization,
depending on the exact value of $Q_0$. $\iso{Ho}{163}$ has been studied
previously in the context of calorimetric $m_\nu$ measurements in
ref.~\cite{DeRujula:1982qt}.
In Fig.~\ref{fig:bb}, we plot $[\Gamma(m_\nu = 0) - \Gamma(m_\nu \neq 0)] /
\Gamma(m_\nu = 0)]$, the relative effect of nonzero $m_\nu$ on the \bb\ or EC
decay rate, as a function of $Q$. We see that even if $Q \sim 100$~eV is
achieved, the effect of $m_\nu = 2$~eV ($0.2$~eV) is only at the level of
$10^{-4}$ ($10^{-6}$). Even if all systematic uncertainties could be reduced to
that level, detecting a deviation from the $m_\nu = 0$ case would still require
the observation of $\text{few} \times 10^{8}$ ($10^{12}$) low-$Q$ decays.  To
complete the experiment within few years of measurement time, this would in
turn require a very large and continuously replenished sample of about
$10^{16}$ ($10^{20}$) stored parent particles\footnote{To arrive at this
estimate, we have taken the known partial lifetime of the EC decay
$\iso{Dy}{159} \rightarrow \iso{Tb^*}{159}$ with $E^* =
363.51$~keV~\cite{Firestone} and have replaced the phase space factor and the
electron wave function by the expressions appropriate for the low-$Q$ decay.},
implying extreme, and possibly prohibitive, requirements on isotope production
and (in the case of ionized parent atoms) storage technology. Part of the
problem is the fact that the nuclear matrix elements for the relevant decay
mode are small. If they were of $\mathcal{O}(1)$, the decay rate would be about
$10^4$ times larger.  Let us emphasize again that decays with larger matrix
elements (or even smaller $Q$ values) may exist, but to identify them, more
precise data on $Q_0$ values and on nuclear excitation levels is needed.

\section{Feasibility of a neutrino mass measurement using low-$Q$ decays}
\label{sec:feas}

In order to exploit the high $m_\nu$-sensitivity of low-$Q$ decays, one has to
overcome several severe technological challenges. We consider the most
important ones to be (A) producing a sufficient number of parent nuclei, (B)
storing them, (C) obtaining an accurate prediction for the decay rate (for \bb\
and EC decays, where no spectral information is available), and (D) counting
the decays. In the following, we discuss some ideas on how these difficulties
might be overcome.

{\bf (A) Producing a sufficient number of parent nuclei.} Most of the isotopes
listed in Table~\ref{tab:isotopes} are unstable, so they would have to be
produced artificially. At future facilities like FAIR at GSI, radioactive beams
with at least $10^8$ -- $10^{10}$ ions/s can be produced~\cite{FAIR-CDR} for
nuclei not too far from stability. For isotopes with half lives of
$\mathcal{O}(\text{days})$, this is in principle sufficient to sustain a sample
of $10^{13}$--$10^{15}$ parent particles, but our discussion above shows that a
competitive neutrino mass measurement would still require an improvement of
several orders of magnitude unless a new, extremely favorable low-$Q$ decay
mode is discovered in the future.  For an experiment using ionized parent
atoms, an additional challenge is to remove ions in other than the desired
charge state in order to avoid decays with larger $Q$ value but identical
experimental signature (i.e.\ identical $\gamma$ and x-ray fingerprint) as the
considered low-$Q$ decay. Due to the different charge-over-mass ratios of
differently charged ions, this should in principle be possible.

{\bf (B) Storing a sufficient number of parent particles.} While an experiment
using neutral atoms (e.g.\ $\iso{Dy}{159}$ and $\iso{Ho}{163}$) can use a
gaseous, liquid, or solid source, a setup using ions requires a trap or a
storage ring. With present technology, it is possible to store a total charge
of $10^8 e$ (corresponding to $10^6$ heavy ions) in a
trap~\cite{Ames:2005a,PodaderaAliseda:2006rx} and $10^{11} e$ in a storage
ring~\cite{KalantarNayestanaki:2009a}. Traps might be pushed to $10^9
e$~\cite{Blaum:PrivComm} in the future, and the planned FAIR facility at GSI
Darmstadt would provide storage rings with a capacity of $10^{12}
e$~\cite{FAIR-CDR}.  As already mentioned, this is still not sufficient to perform
a low-$Q$ $\beta$ or EC decay experiment using ionized parent atoms
competitive to KATRIN unless new decay modes with $Q < 1$~eV and large nuclear
matrix element are discovered. This implies that, from the present perspective,
decays of neutral atoms look more promising.

{\bf (C) Predicting the decay rate $\Gamma$ for \bb\ or EC decay.} The main
unknowns in the computation of $\Gamma$ are the nuclear matrix element, the
nuclear mass difference, and the electron wave functions. To avoid the
uncertainty in the matrix element, we propose to study not only the low-$Q$
decay, but also a large-$Q$ (i.e.\ high rate, but small $m_\nu$-dependence)
decay into the same nuclear final state to measure the nuclear matrix element.
The mass difference $W_0$ between the parent and daughter nuclei can be
measured using ion trap mass spectrometry. This technique currently provides an
impressive relative accuracy of
$\mathcal{O}(10^{-11})$~\cite{KlausHABIL,Blaum:2009eu}, but for our purposes,
this would still have to be increased by more than one order of magnitude to
make the uncertainty in $W_0$ smaller than the effect of the neutrino mass. The
electron wave functions entering in $\Gamma$ cannot be measured directly and
have to be predicted by solving the multi-particle Dirac equation. The
uncertainties of these predictions must be smaller than the expected effect of
$m_\nu$, but considering that many atomic x-ray spectra can be predicted to an
accuracy below one per mille~\cite{Deslattes:2003zz}, this could be feasible.
To minimize the theoretical errors, one could `calibrate' the numerical
computation using experimental x-ray spectra, ionization energies, and other
atomic physics data for the considered isotope.

{\bf (D) Counting the number of decays.} For \bb\ and EC decay, the only
observable sensitive to $m_\nu$ is the decay rate into the low-$Q$ channel. To
measure it, and to reject concurrent large-$Q$ decay modes, we propose to
detect characteristic gamma or x-ray photons accompanying the decay.  The main
requirements for the photon detector are good solid angle coverage, high energy
resolution, and efficient suppression of backgrounds from cosmic ray
interaction products and radioactive impurities. To date, the best $\gamma$
detectors --- employing extremely radiopure materials, active and passive
shielding, and several meters of rock overburden --- achieve background rates
$\lesssim 10^3\ {\rm keV}^{-1} {\rm yr}^{-1}$ and an energy resolution around
1~keV~\cite{Heusser:PrivComm}.  If the considered low-$Q$ decay is accompanied
by several photons, much better background suppression will be possible if the
coincidence technique is used.  Therefore, we estimate that backgrounds can be
brought under control.

\section{Conclusions}
\label{sec:conc}

In this paper, we have discussed how continuum $\beta$, bound state $\beta$,
and electron capture decays with extremely small $Q$ values ($\ll 1$~keV) can
be realized and how they could be used to measure the absolute neutrino mass
$m_\nu$. To achieve sufficiently low $Q$ values, i.e.\ sufficiently high
sensitivity to $m_\nu$, we have proposed to consider decays into excited
nuclear daughter states, and, if necessary, to partially ionize the atoms to
tune the electronic contribution to $Q$. We have discussed the technological
challenges that would have to be overcome in such an experiment, including
production and storage of a large number of radioactive atoms or ions,
obtaining accurate predictions for the decay rate as a function of $m_\nu$, and
counting the number of decays. We have found that the most promising decays to
date are $\iso{Dy}{159} \rightarrow \iso{Tb^*}{159}$ and $\iso{Ho}{163}
\rightarrow \iso{Dy}{163}$ because, depending on the exact values of the
respective atomic mass differences $Q_0$, they may have low-$Q$ EC decay modes
even when neutral. Experiments using ions are much more challenging due to the
large number of particles that must be stored.  As a next step, it is crucial
to measure precisely $Q_0$ for the isotopes listed in Table~\ref{tab:isotopes}
in order to determine how small $Q$ can be made for them. Also, more precise
data on nuclear excitation spectra throughout the chart of nuclides is
desirable in order to identify further candidates for low $Q$ decays.

\section*{Acknowledgments}
\label{sec:Ack}

It is a pleasure to thank K.~Blaum, F.~Bosch, G.~Heusser, B.~Kayser,
M.~Lindner, Yu.~Litvinov, Yu.~Novikov, and M.~Weber for useful and inspiring
discussions.  This work has been supported by the DFG-SFB TR~27 `Neutrinos and
beyond'.  Fermilab is operated by Fermi Research Alliance, LLC under Contract
No.~DE-AC02-07CH11359 with the US Department of Energy.


\end{document}